\documentclass{vgtc}
\ifpdf
  \pdfoutput=1\relax
  \usepackage{graphicx}
  \DeclareGraphicsExtensions{.pdf,.png,.jpg,.jpeg}
\else
  \usepackage{graphicx}
  \DeclareGraphicsExtensions{.eps}
\fi
\graphicspath{{figures/}{pictures/}{images/}{./}}
\usepackage[abbreviations]{glossaries-extra}
\usepackage{microtype}
\usepackage{pifont}
\usepackage{textcomp}
\PassOptionsToPackage{warn}{textcomp}
\usepackage{mathptmx}
\usepackage{times}
\usepackage{cite}
\usepackage{tabu}
\usepackage{tabularx}
\usepackage{multirow}
\usepackage{booktabs}
\usepackage[table]{xcolor}
\usepackage{array, makecell}
\usepackage{hyperref}
\usepackage{cleveref}
\usepackage{xspace}
\usepackage{savesym}
\usepackage[final]{pdfpages}

\usepackage{xcolor}
\definecolor{maroon}{cmyk}{0, 0.87, 0.68, 0.32}
\definecolor{halfgray}{gray}{0.55}
\definecolor{ipython_frame}{RGB}{207, 207, 207}
\definecolor{ipython_bg}{RGB}{247, 247, 247}
\definecolor{ipython_red}{RGB}{186, 33, 33}
\definecolor{ipython_green}{RGB}{0, 128, 0}
\definecolor{ipython_cyan}{RGB}{64, 128, 128}
\definecolor{ipython_purple}{RGB}{170, 34, 255}

\usepackage{listings}
\lstdefinelanguage{iPython}{
    morekeywords={access,and,break,class,continue,def,del,elif,else,except,exec,finally,for,from,global,if,import,in,is,lambda,not,or,pass,print,raise,return,try,while},%
    morekeywords=[2]{abs,all,any,basestring,bin,bool,bytearray,callable,chr,classmethod,cmp,compile,complex,delattr,dict,dir,divmod,enumerate,eval,execfile,file,filter,float,format,frozenset,getattr,globals,hasattr,hash,help,hex,id,input,int,isinstance,issubclass,iter,len,list,locals,long,map,max,memoryview,min,next,object,oct,open,ord,pow,property,range,raw_input,reduce,reload,repr,reversed,round,set,setattr,slice,sorted,staticmethod,str,sum,super,tuple,type,unichr,unicode,vars,xrange,zip,apply,buffer,coerce,intern},%
    sensitive=true,%
    morecomment=[l]\#,%
    morestring=[b]',%
    morestring=[b]",%
    morestring=[s]{'''}{'''},
    morestring=[s]{"""}{"""},
    morestring=[s]{r'}{'},
    morestring=[s]{r"}{"},%
    morestring=[s]{r'''}{'''},%
    morestring=[s]{r"""}{"""},%
    morestring=[s]{u'}{'},
    morestring=[s]{u"}{"},%
    morestring=[s]{u'''}{'''},%
    morestring=[s]{u"""}{"""},%
    literate=
    {á}{{\'a}}1 {é}{{\'e}}1 {í}{{\'i}}1 {ó}{{\'o}}1 {ú}{{\'u}}1
    {Á}{{\'A}}1 {É}{{\'E}}1 {Í}{{\'I}}1 {Ó}{{\'O}}1 {Ú}{{\'U}}1
    {à}{{\`a}}1 {è}{{\`e}}1 {ì}{{\`i}}1 {ò}{{\`o}}1 {ù}{{\`u}}1
    {À}{{\`A}}1 {È}{{\'E}}1 {Ì}{{\`I}}1 {Ò}{{\`O}}1 {Ù}{{\`U}}1
    {ä}{{\"a}}1 {ë}{{\"e}}1 {ï}{{\"i}}1 {ö}{{\"o}}1 {ü}{{\"u}}1
    {Ä}{{\"A}}1 {Ë}{{\"E}}1 {Ï}{{\"I}}1 {Ö}{{\"O}}1 {Ü}{{\"U}}1
    {â}{{\^a}}1 {ê}{{\^e}}1 {î}{{\^i}}1 {ô}{{\^o}}1 {û}{{\^u}}1
    {Â}{{\^A}}1 {Ê}{{\^E}}1 {Î}{{\^I}}1 {Ô}{{\^O}}1 {Û}{{\^U}}1
    {œ}{{\oe}}1 {Œ}{{\OE}}1 {æ}{{\ae}}1 {Æ}{{\AE}}1 {ß}{{\ss}}1
    {ç}{{\c c}}1 {Ç}{{\c C}}1 {ø}{{\o}}1 {å}{{\r a}}1 {Å}{{\r A}}1
    {€}{{\EUR}}1 {£}{{\pounds}}1,
    literate=
    *{+}{{{\color{ipython_purple}+}}}1
    {-}{{{\color{ipython_purple}-}}}1
    {*}{{{\color{ipython_purple}$^\ast$}}}1
    {/}{{{\color{ipython_purple}/}}}1
    {^}{{{\color{ipython_purple}\^{}}}}1
    {?}{{{\color{ipython_purple}?}}}1
    {!}{{{\color{ipython_purple}!}}}1
    {\%}{{{\color{ipython_purple}\%}}}1
    {<}{{{\color{ipython_purple}<}}}1
    {>}{{{\color{ipython_purple}>}}}1
    {|}{{{\color{ipython_purple}|}}}1
    {\&}{{{\color{ipython_purple}\&}}}1
    {~}{{{\color{ipython_purple}~}}}1
    {==}{{{\color{ipython_purple}==}}}2
    {<=}{{{\color{ipython_purple}<=}}}2
    {>=}{{{\color{ipython_purple}>=}}}2
    {+=}{{{+=}}}2
    {-=}{{{-=}}}2
    {*=}{{{$^\ast$=}}}2
    {/=}{{{/=}}}2,
    commentstyle=\color{ipython_cyan}\ttfamily,
    stringstyle=\color{ipython_red}\ttfamily,
    keepspaces=true,
    showspaces=false,
    showstringspaces=false,
    rulecolor=\color{ipython_frame},
    frame=single,
    frameround={t}{t}{t}{t},
    framexleftmargin=0mm,
    numbers=left,
    numberstyle=\tiny\color{halfgray},
    backgroundcolor=\color{ipython_bg},
    basicstyle=\scriptsize\ttfamily,
    keywordstyle=\color{ipython_green}\ttfamily,
    escapechar=\¢,escapebegin=\color{ipython_green},
}

\savesymbol{spacing}

\definecolor{Orange}{rgb}{1.0, 0.5, 0.0}
\definecolor{DarkGreen}{rgb}{0, 0.5, 0.0}
\definecolor{Purple}{rgb}{0.7, 0.0, 0.7}
\definecolor{Blue}{rgb}{0.2, 0.2, 0.8}
\definecolor{Red}{rgb}{1.0, 0.0, 0.0}
\definecolor{Brown}{rgb}{0.7, 0.4, 0.1}
\definecolor{Blue}{rgb}{0, 0, 1.}
\definecolor{Green}{rgb}{0., 0.6, 0.}
\definecolor{Custom}{rgb}{0.3, 0.1, 0.2}
\definecolor{Yellow}{rgb}{0.9, 0.7, 0.0}
\definecolor{Purple}{rgb}{0.9, 0.1, 0.8}

\newcommand{\etal}{~et al.\@\xspace}

\newcommand{\eg}{e.g.\@\xspace}

\newcommand{\ie}{i.e.\@\xspace}

\newcommand{\codeoptimization}{\textcolor{blue}{\href{https://github.com/complight/learned_prescription}{$GitHub:complight/learned\_prescription$}}\@\xspace}
\newcommand{\codemodel}{\textcolor{blue}{\href{https://github.com/complight/learned_prescription_model}{$GitHub:complight/learned\_prescription\_model$}}\@\xspace}

\newabbreviation{HVS}{HVS}{Human Visual System}
\newabbreviation{AR}{AR}{Augmented Reality}
\newabbreviation{VR}{VR}{Virtual Reality}
\newabbreviation{FoV}{FoV}{Field Of View}
\newabbreviation{HOE}{HOE}{Holographic Optical Element}
\newabbreviation{3D}{3D}{Three-Dimensional}
\newabbreviation{CNN}{CNN}{Convolutional Neural Network}
\newabbreviation{PSF}{PSF}{Point-Spread Function}
\newabbreviation{MLP}{MLP}{Multilayer Perceptron}

\global\long\def\CNN{\gls{CNN}\xspace}
\global\long\def\HVS{\gls{HVS}\xspace}
\global\long\def\3D{\gls{3D}\xspace}
\global\long\def\AR{\gls{AR}\xspace}
\global\long\def\VR{\gls{VR}\xspace}

\global\long\def\PSF{\gls{PSF}\xspace}
\global\long\def\MLP{\gls{MLP}\xspace}

\newcommand{\refSec}[1]{Sec.~\ref{sec:#1}}
\newcommand{\refFig}[1]{Fig.~\ref{fig:#1}}
\newcommand{\refEq}[1]{Eq.~\ref{eq:#1}}
\newcommand{\refTbl}[1]{Tbl.~\ref{tbl:#1}}

\DeclareMathOperator*{\argmin}{{\arg\!\min}}
\newcolumntype{P}[1]{>{\centering\raggedright\arraybackslash}p{#1}}
\newcolumntype{M}[1]{>{\centering\arraybackslash}m{#1}}

\onlineid{1215}
\vgtccategory{Research}
\vgtcinsertpkg

\title{ChromaCorrect: Prescription Correction in Virtual Reality Headsets through Perceptual Guidance}

\author{Ahmet Güzel\thanks{e-mail: od20ahg@leeds.ac.uk}\\
        \scriptsize University of Leeds
\and Jeanne Beyazian\thanks{e-mail: jeanne.beyazian.21@ucl.ac.uk}\\
        \scriptsize University College London
\and Praneeth Chakravarthula\thanks{e-mail: praneethc@princenton.edu}\\
     \parbox{1.4in}{\scriptsize \centering UNC Chapel Hill \\ Princenton University}
\and Kaan Akşit\thanks{e-mail: k.aksit@ucl.ac.uk}\\
	\scriptsize University College London}

\teaser{
\centering
\includegraphics[width=1.0\columnwidth]{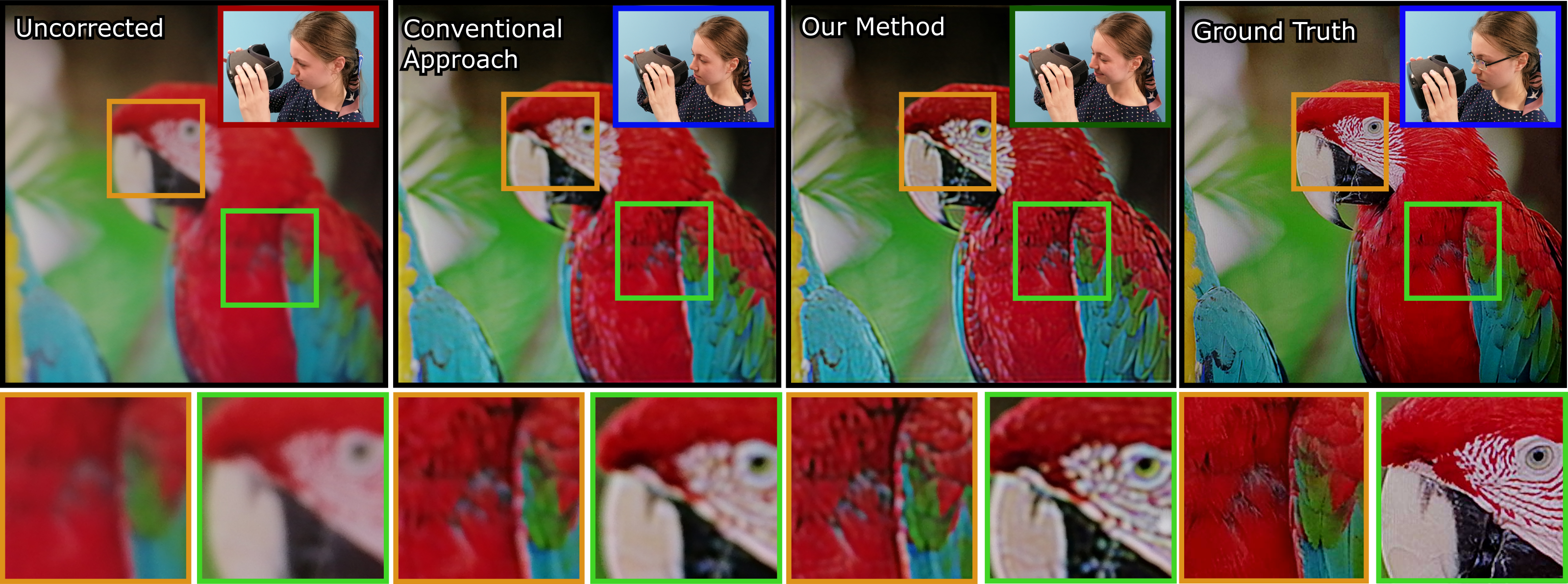}
\caption{
Perceptual Guidance in Prescription Correction.
We provide a differentiable perception model for optimizing images that compensate for user prescription and improve the perceived contrast and color in images.
Here, we show images as captured by a camera on a prototype Virtual Reality headset where the images are intentionally defocused to mimic an eye with common refractive errors.
Without any prescription correction, the perceived images appear blurry due to defocus caused by refractive errors (first column).
The second column captures the performance of the conventional algorithmic approach to prescription correction \cite{montalto2015total} for the same refractive error. 
Our proposed computational appraoch to algorithmic prescription compensation improves the perceived images, both in color and contrast, as can be seen in the third column.
For reference, we provide a ground truth photograph focused at the display plane as in the fourth column, resembling what a user would see with their prescription lenses incorporated into the virtual reality headset.
Source image is from Rich Franzen~\cite{franzen1999kodak}.
}
\label{fig:teaser}
}

\abstract{
A large portion of today's world population suffer from vision impairments and wear prescription eyeglasses.
However, eyeglasses causes additional bulk and discomfort when used with augmented and virtual reality headsets, thereby negatively impacting the viewer's visual experience.
In this work, we remedy the usage of prescription eyeglasses in Virtual Reality (VR) headsets by shifting the optical complexity completely into software and propose a prescription-aware rendering approach for providing sharper and immersive VR imagery.
To this end, we develop a differentiable display and visual perception model encapsulating display-specific parameters, color and visual acuity of human visual system and the user-specific refractive errors.
Using this differentiable visual perception model, we optimize the rendered imagery in the display using stochastic gradient-descent solvers.
This way, we provide prescription glasses-free sharper images for a person with vision impairments. 
We evaluate our approach on various displays, including desktops and VR headsets, and show significant quality and contrast improvements for users with vision impairments.
}

\CCScatlist{
  \CCScatTwelve{Human-centered computing}{Visu\-al\-iza\-tion}{Visu\-al\-iza\-tion techniques}{Treemaps};
  \CCScatTwelve{Human-centered computing}{Visu\-al\-iza\-tion}{Visualization design and evaluation methods}{}
}

\begin{document}

\firstsection{Introduction}
\label{sec:introduction}
\maketitle

\VR headsets are becoming increasingly popular amongst consumers.
This encouraged researchers to conceptualize and build technologies that would enable fully immersive remote experiences  \cite{Orlosky2021}.
However, most of the recent developments overlook the prevalence of refractive vision problems such as myopia, hyperopia, or astigmatism among potential VR users especially older than 40 years old, which is at least 23.9\%, 8.4\% and 33\% of population, respectively \cite{AAO:2019}.
Moreover, while the current near-eye display research is focused on miniaturization of the headset to eyeglasses form-factor \cite{Maimone2020Jul, kim2022holographic}, wearing prescription glasses under a \VR headset causes uncomfortable viewing experiences that break the feeling of immersion.

Hardware-driven approaches to prescription correction \cite{wu2020prescription, Kim2019matching, chakravarthula2018focusar} may lead to \VR headsets that are bulkier and expensive while necessitating upgrading components with new devices.
On the other hand, algorithmic approaches to prescription correction enable tackling the prescription issue without the need for specialized components and with the benefit of software updates \cite{montalto2015total}.

Our work offers a new perceptually-guided algorithmic approach to prescription correction, while eliminating the need of corrective lenses.
To this end, we first study the low-level workings of the Human Visual System (HVS), \ie, how different types of cone cells respond to various wavelengths of light.
We then model the display's specific light spectrum (\eg subpixels emitting various wavelengths) and the associated response of cone cells on the retina.
Hence, we build an end-to-end differentiable perception model that helps us to simulate how a user with a \PSF model with Zernike polynomials \cite{lakshminarayanan2011zernike} perceives images on a specific display.
Finally, our end-to-end perception framework enables optimizing the display rendering to produce an in-focus image for a user with vision impairments.
Specifically, our work makes the following contributions:

\begin{itemize}

\item \textbf{Perceptually guided Prescription Correction.}
We incorporate the display specific color perception and \PSF of a user into a new differentiable model to ensure that the optimized image's contrast and color characteristics are distinctly enhanced in visual perception.

\item \textbf{Learned Prescription Correction.}
We train a \CNN to estimate optimal images for prescription correction, enabling prescription correction at interactive rates.

\item \textbf{Evaluation on Actual Displays.}
We analyze our findings beyond simulations.
Thus, we evaluate our approach to \VR headsets and conventional displays and demonstrate real-life use cases.

\end{itemize}

\section{Related Work}
\label{sec:related_work}
Researchers have previously attempted to compensate for refractive vision problems for glasses-free experience in displays. 
We summarize most relevant papers here in \refTbl{comparison_table}.

\renewcommand{\arraystretch}{1.2} 
\begin{table}[!ht]
\caption{
Comparison of prescription correction techniques.
Many of the solutions for prescription correction either fail to provide good image quality or require bulky hardware components affecting user comfort negatively.
We take an algorithmic approach utilizing an accurate perception model of the human visual system, leading to improved image quality and real-time image generation.
SW refers to Software while HW refers to Hardware in this table.
}
\label{tbl:comparison_table}
\noindent
  \begin{tabularx}{0.48\textwidth}{P{2.0cm}P{0.6cm}P{1.3cm}P{0.6cm}P{0.6cm}P{1cm}}
    \toprule
	  \multirow{1}{*}{\thead{Name}} & \multirow{1}{*}{\thead{Method}} & \multirow{1}{*}{\thead{Perceptual \\ Guidance}} & \multirow{1}{*}{\thead{Realtime}} & \multirow{1}{*}{\thead{Image \\ Quality}} & \multirow{1}{*}{\thead{Display \\ Type}} \\
    \\
    \midrule
    Multi-domain
    \cite{alonso2006multi-domain} & \cellcolor{green!25}SW & \cellcolor{yellow!25}Preliminary& \cellcolor{red!25}No & \cellcolor{red!25}Poor & Desktop \\

    Constrained Total Variation \textsuperscript{*}
    \cite{montalto2015total} & \cellcolor{green!25}SW & \cellcolor{yellow!25}Preliminary  & \cellcolor{red!25}No  & \cellcolor{red!25}Poor & Desktop \\

    Tone Mapping
    \cite{ye2018content} & \cellcolor{green!25}SW & \cellcolor{yellow!25}Preliminary & \cellcolor{red!25}No  & \cellcolor{red!25}Poor & Desktop \\

    Network
    \cite{tanaka2021image} & \cellcolor{green!25}SW & \cellcolor{red!25}No  & \cellcolor{red!25}No  & \cellcolor{red!25}Poor & Desktop \\

    Vision Enhancement
    \cite{Itoh2015Mar} & \cellcolor{green!25}SW & \cellcolor{red!25}No & \cellcolor{red!25}No  & \cellcolor{red!25}Poor & AR \\

    SharpView
    \cite{oshima2016sharpview} & \cellcolor{green!25}SW & \cellcolor{red!25}No & \cellcolor{red!25}No  & \cellcolor{red!25}Poor & AR \\

    FocusAR
    \cite{chakravarthula2018focusar} & \cellcolor{red!25}HW & \cellcolor{red!25}No & \cellcolor{green!25}Yes  & \cellcolor{green!25} Good & AR \\

    Autofocals
    \cite{padmanaban2019autofocals} & \cellcolor{red!25}HW & \cellcolor{red!25}No & \cellcolor{green!25}Yes & \cellcolor{green!25} Good & AR \\

    Phase Modulated
    \cite{Itoh2021Computational} & \cellcolor{red!25}HW & \cellcolor{red!25}No  & \cellcolor{green!25}Yes & \cellcolor{green!25} Good  & AR \\

    RectifEye
    \cite{10.1145/2996376.2996382} & \cellcolor{red!25}HW & \cellcolor{red!25}No & \cellcolor{green!25}Yes & \cellcolor{green!25} Good  & VR \\

    Alvarez Lenses
    \cite{10.1117/12.2318397} & \cellcolor{red!25}HW & \cellcolor{red!25}No & \cellcolor{green!25}Yes & \cellcolor{green!25} Good  & VR \\

    Software
    \cite{xu2018software} & \cellcolor{green!25}SW & \cellcolor{red!25}No & \cellcolor{green!25}Yes  & \cellcolor{red!25} Poor & VR \\

    Ours & \cellcolor{green!25}SW & \cellcolor{green!25}Yes & \cellcolor{green!25}Yes & \cellcolor{yellow!25} Fair & VR \\
    \bottomrule
    \multicolumn{6}{p{\dimexpr\linewidth-2\tabcolsep\relax}}{\textsuperscript{*}\footnotesize{This technique is refered to as the conventional method throughout the paper.}}
  \end{tabularx}
\end{table}
\renewcommand{\arraystretch}{1.0} 

\paragraph{Programmable Prescription Lenses.}
Utilizing focus-tunable lenses that may be adjusted to the user's prescription is a common technique, especially in displays such as \VR headsets where the users view a display through magnifying lenses \cite{10.1117/12.2318397, 10.1145/2996376.2996382, chakravarthula2018focusar, padmanaban2019autofocals}.
An alternative to these approaches, phase-only spatial light modulators, could also be used to form a programmable prescription correction lens \cite{Itoh2021Computational}.
Beyond requiring customized hardware, these techniques would also require eye-tracking and depth sensor data of a scene to operate, leading to more demands in hardware.

\paragraph{Computational Displays.}
Altering the display hardware and image acquisition technologies could help with prescription correction \cite{koulieris2019near}.
Huang\etal \cite{huang2012correcting} address extreme contrast loss and ringing artifacts in algorithmic correction techniques by utilizing a stack of semi-transparent, light-emitting layers for LCDs.
Wu and Kim \cite{Wu:20} embed free-form image combiners inside prescription lenses to create customizable \AR displays.
Pamplona\etal \cite{Pamplona2012Jul} implements 4D light field displays to move the solution to a higher-dimensional (light field) space, where the inverse problem is well-posed.
To overcome this limitation in resolutions in Pamplona's work \cite{Pamplona2012Jul}, Huang\etal \cite{Huang2014Jul} propose a 4D prefiltering algorithm that can provide higher contrasts and resolutions.
The described approach \cite{Pamplona2012Jul} has a significant drawback, namely that the \PSF of an eye with refractive errors is typically a low-pass filter and, as such, irrevocably cancels higher frequencies from the original image.
Moreover, holographic vision correction \cite{Kim:21,chakravarthula2021gaze} is superior to conventional approaches, including light field displays.
Curious readers could consult to survey by Aydınoğlu\etal \cite{aydindougan2021applications} for more on these holographic displays.

\paragraph{Algorithmic Prescription Correction.}
Refractive vision impairments of the eye are commonly modeled by constructing a \PSF that represents how the eye as an optical system transmits a point on the object to a point on the retina.
The spatially varying \PSF is convolved with the image of the object to produce the image formed on the retina.
Performing the inverse operation, \ie, deconvolving the image with the retinal \PSF, could help produce an image that forms clearly on the retina when observed.
Alonso\etal \cite{alonso2005verification} verifies the possibility of such an image correction technique by constructing a simple artificial eye and comparing the image it forms when viewing a standard and a corrected image.
They also propose an ad-hoc solution to mitigate contrast loss and ``ripples'' or ringing artifacts \cite{alonso2006multi-domain}.
Monalto\etal \cite{montalto2015total} present constrained total variation to decrease ringing artifacts in the corrected image while sharpening the image's edges, thereby producing an image with high contrast along sharp edges.
Ye\etal \cite{ye2018content} focus on finding a ringing-free image with higher contrast in locations important to \HVS, while tolerating more blurriness elsewhere.
Tanaka\etal \cite{tanaka2021image} uses a CNN-based pipeline for prescription correction along with Zernike-based visual aberration modeling.
Li\etal \cite{li2021universal} feed an aberrated image and a map of a \PSF for multiple subregions, to account for spatially variant aberrations into a deep neural network and train it for image correction on a variety of lenses.
Similar image correction techniques have been applied to \VR headsets. 
Itoh\etal \cite{Itoh2015Mar} corrects the defocus aberration for optical see-through headsets by overlaying a compensated image in the user's view.
Xu\etal \cite{xu2018software} use gradient-based priors to achieve realtime visual aberration correction for VR HMDs.
Oshima\etal \cite{oshima2016sharpview} describe realtime defocus correction for optical see-through HMDs, which is caused by focal rivalry: the simultaneous viewing of real and virtual content.

Perceptual considerations in displays and graphics systems are becoming commonplace in relevant research branches (\eg consult our supplementary for perceptual considerations in graphics systems).
The surveyed research work does not provide a complete model of \HVS in their solutions, leading to either poor image quality or demanding hardware.
We believe our work resembles the first attempt to enhance algorithmic solutions in the literature by bridging the gap between perceptual modeling and prescription correction.

\begin{figure*}[!ht]
\begin{centering}
\includegraphics[width=1.9\columnwidth]{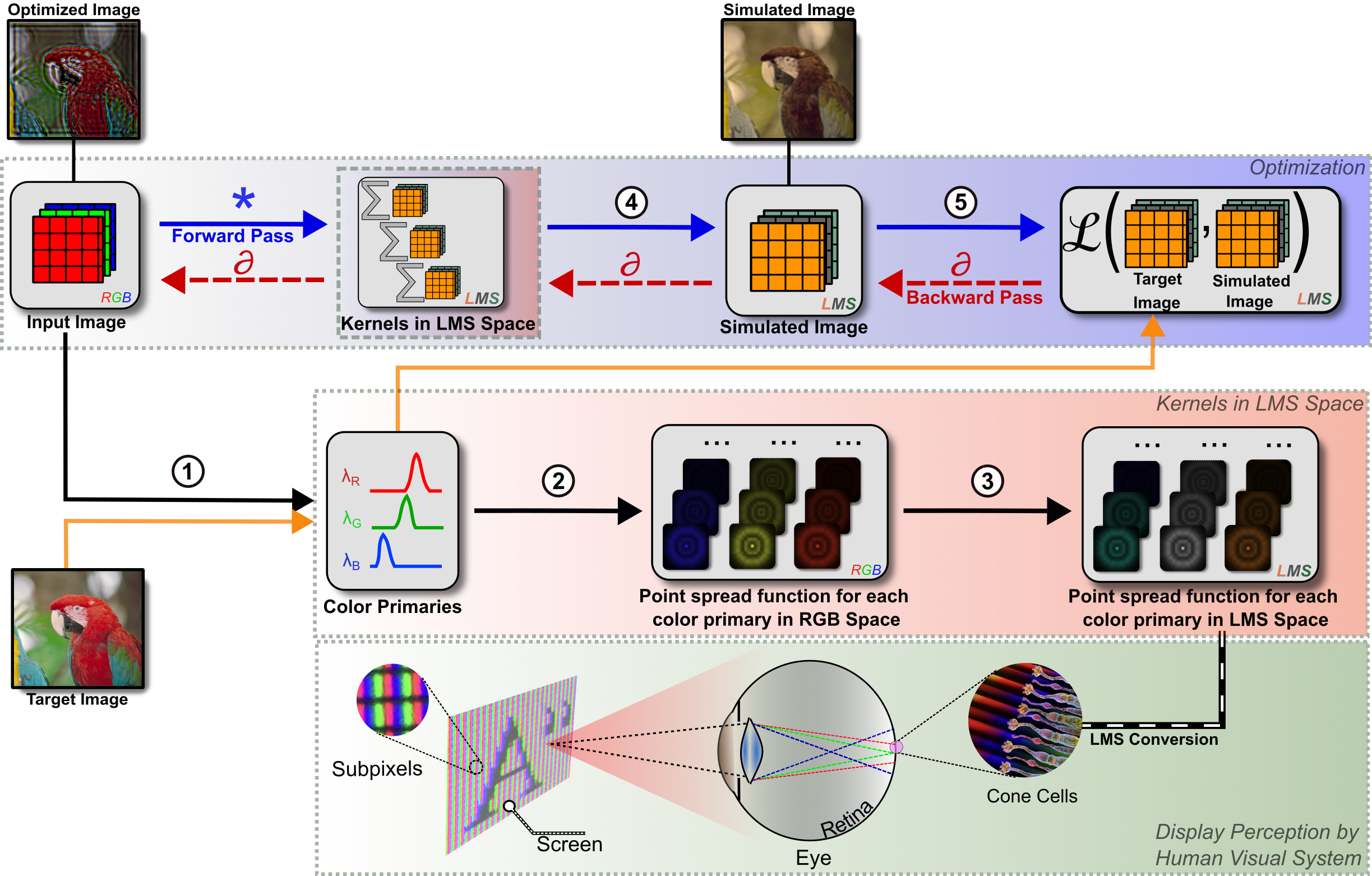}
\caption{
Prescription correction using a perceptually guided computational model and a differentiable optimization pipeline.
(1) A screen with color primaries (RGB) displays an input image.
(2) A viewer's eye images the displayed image onto the retina with a unique Point Spread Function (PSF) describing the optical aberrations of that person's eye.
(3) Retinal cells convert the aberrated RGB image to a trichromat sensation, also known as Long-Medium-Short (LMS) cone perception \cite{STOCKMAN20001711}.
(4) Our optimization pipeline relies on the perceptually guided model described in previous steps (1-3).
Thus, the optimization pipeline converts a given RGB image to LMS space at each optimization step while accounting for the PSFs of a viewer modelled using Zernike polynomials.
(5) Our loss function penalizes the simulated image derived from the perceptually guided model against a target image in LMS space.
Finally, our differentiable optimization pipeline identifies proper input RGB images using a Stochastic Gradient Descent solver \cite{paszke2017automatic}.
}
\label{fig:system_model}
\end{centering}
\end{figure*}

\section{Perceptually Guided Prescription Correction}
\label{sec:method}
We introduce a differentiable framework for modeling the display and human visual perception, encapsulating display-specific parameters, color and visual acuity of human visual system and the user-specific refractive errors. 
Our framework allows for optimizing prescription compensated rendered imagery on standard displays using a gradient-based policy with novel display-specific perceptually guided loss functions (Section \ref{subsec:display_specific_perceptually_guided_loss}).
We rely on Zernike polynomials (Section \ref{subsec:defining_presecription_models}) for describing user-specific retinal point spread functions \cite{chakravarthula2021gaze} within the forward model to represent optical aberrations in the \HVS (Section \ref{subsec:optimizing_images_for_prescription_correction}).
On overview of our entire display-visual perception model and the optimization process is depicted in \refFig{system_model}.

\subsection{Modeling Display-specific Visual Perception}
\label{subsec:display_specific_perceptually_guided_loss}
We characterize our target display and device a computational model to transform the displayed imagery on the target display into imagery as perceived by the \HVS.

\paragraph{Characterizing target display.}
A given display has three types of emission spectra, $\lambda_R, \lambda_G, \lambda_B$, for their red, green, and blue channel pixels, respectively. 
As these emission spectra vary for each display system, we calibrate the spectra using a spectrometer by measuring the spectral bands of the target display at various pixel levels.
More details on the spectra measurement and display calibration process are discussed in the Supplementary Material.
We then fit a proxy function to determine the display color primaries from the spectral measurements. 
While a simple Gaussian mixture model with weighted sum of Gaussians can be used for such a proxy color primary function, we learn this function using a multi-layer perceptron network that act as general function approximator. 
Implementation of this proxy function fitting can be found in (See $odak.learn.tools.multi\_layer\_perceptron()$ in \cite{aksit_kaan_2022_6528486}).
Once we fit a proxy function for the color primaries, we utilize it to investigate the color perception responses of the \HVS.

\paragraph{Converting color primaries to perceived colors.}
Human retinal cells can be broadly classified into rods and cones.
Cone cells, which are primarily responsible for color perception in the \HVS, are of three different subtypes: Short (S), Medium (M), and Long (L) cells.
Each of them differs in its sensitivity to different wavelengths of light. 
Please refer to our Supplementary Material for a detailed discussion this.
The L, M, and S cones reduce wavelengths of incoming light into trichromat values by integrating them over their response functions~\cite{Wuerger2022Colour}.
Note that perception in \HVS is contrary to modeling general camera or display response where red, green and blue wavelengths are independently measured on the camera sensor or the human retina.
The following steps show how to convert an input color image displayed on a target display to the corresponding cone response:
\begin{equation}
\label{eq:input_image_to_perception_1st}
\begin{bmatrix}
I_{L}\\
I_{M}\\
I_{S}\\
\end{bmatrix}
=
\begin{bmatrix}
L_R& L_G& L_B\\
M_R& M_G& M_B\\
S_R& S_G& S_B\\
\end{bmatrix}
\begin{bmatrix}
I_{R}\\
I_{G}\\
I_{B}\\
\end{bmatrix}
,
\end{equation}
where $I_{R}$, $I_{G}$, $I_{B}$ represents red, green and blue pixel values of an input image, and $I_{L}$, $I_{M}$, $I_{S}$ represents L, M and S cone activation values for each pixel of the displayed image.
From the generalized formula above, we provide a sample conversion for $L_R$ as in the following equation,
\begin{equation}
\sum_{\lambda_R} \lambda_L \lambda_R = L_R
,
\end{equation}
where $\lambda_L$ represents L cone sensitivity function, $\lambda_R$ represents red pixel emission spectrum function for a targeted display, and $L_R$ represents L cone output for the displayed red pixel.
Similarly, L cone sensitivity functions for green and blue pixel emissions can be computed.
Thus, L, M and S cone sensitivity functions can be computed for the three different subpixel emissions.
After computing the cone sensitivity functions, we apply the conversion from the color opponency
model proposed by Schmidt\etal \cite{Schmidt2014NeurobiologicalHO} to represent a complete perception model,
\begin{equation}
\label{eq:input_image_to_perception_2nd}
\begin{bmatrix}
I_{(M+S)-L}\\
I_{(L+S)-M}\\
I_{\overline{(L+M+S)}}\\
\end{bmatrix}
=
\begin{bmatrix}
(I_{M} + I_{S}) - I_{L}\\
(I_{L} + I_{S}) - I_{M}\\
\overline{(I_{L}, I_{M}, I_{S}})\\
\end{bmatrix}
,
\end{equation}
where $I_{(M+S)-L}$, $I_{(L+S)-M}$, $I_{\overline{(L,M,S)}}$ represents the three channels of the image sensed in the color-opponency space.

\subsection{Computing Point Spread Functions from Color Primaries}
\label{subsec:defining_presecription_models}
The point spread function for the HVS with visual aberrations can be defined over several wavelengths of light (see Supplementary Material for equations).
Therefore, we can sample a set of wavelengths from each color primary, calculate PSFs for each and use a weighted sum of the PSFs to obtain a single, combined PSF for each color primary,
\begin{equation}
PSF(x, y, c) = \sum_{\lambda_c} w_{\lambda_{c_i}} PSF(x, y, {\lambda}_{c_i})
\label{eq:psf_color_primary}
\end{equation}
where $c$ represents a particular color primary, $PSF(x, y, c)$ is the PSF for a particular color primary, $PSF(x, y, {\lambda}_{c_i})$ the PSF for a sampled wavelength in the color primary and $w_{\lambda_{c_i}}$ is the weight for that sampled wavelength.
The above PSF kernel can be utilized in RGB, or color opponency spaces, depending on designers choices.
In our method, we introduce color opponency based PSF formulation (perceptually guided) to improve the perceptual characteristics (contrast, quality) of the retinal image.
\refEq{psf_color_primary} is extended to formulate LMS based kernel,
\begin{equation}
\label{eq:psf_lms_lambda}
  PSF_{lms}(x, y, {\lambda}_{c_i}) = A *  PSF(x, y, {\lambda}_{c_i})
\end{equation}
\begin{equation}
\label{eq:psf_lms_color_primary}
  PSF_{lms}(x, y, c) = \sum_{\lambda_c} w_{\lambda_{c_i}} PSF_{lms}(x, y, {\lambda}_{c_i})
\end{equation}
where $A$ is the conversion matrix defined in \refEq{input_image_to_perception_1st}, $PSF_{lms}(x, y, c)$ is the PSF for a particular color primary with LMS components.
Similarly, we modelled the digital camera color primary decoding by using measurements from the display and captured images from the digital camera.
In this way, we are able to use digital camera captured images to represent our work in this paper.
In the \refEq{psf_lms_lambda} and \refEq{psf_lms_color_primary}, $PSF_{lms}$ is represented for both the HVS and digital camera RGB decoding.
We can now compute the retinal image $r(x, y, c)$ in the LMS space, by convolving $PSF_{lms}$ with the input image $s(x, y, c)$,
\begin{equation}
\label{eq:retinal_image}
r(x, y, c) = PSF_{lms}(x, y, c) * s(x, y, c).
\end{equation}

\subsection{Optimizing Images for Prescription Correction}
\label{subsec:optimizing_images_for_prescription_correction}
In the final step, we aim to optimize an image which, after passing through the eye's optical system
(modelled as a convolution in \refEq{retinal_image}), is intended to produce a retinal image that is as close as possible to the ground truth image.
This is done by solving the optimization problem,
\begin{equation}
\label{eq:general_optimization}
s' \leftarrow \underset{s\not\in\emptyset}\argmin~\mathcal{L}(PSF*s, t)
\end{equation}
where $t$ is the the ground truth image and $s'$ is the input image optimized for a user's eye, $PSF$ is kernel defined in \refEq{psf_color_primary}.
In our method, we reformulate \refEq{general_optimization} to incorporate color opponency space optimization,
\begin{equation}
s' \leftarrow \underset{s\not\in\emptyset}\argmin~\mathcal{L}(PSF_{lms}*s, t_{lms})
\end{equation}
where $t_{lms}$ is the the ground truth image in LMS space and $s'$ is the input image optimized for a user's eye, $PSF_{lms}$ is kernel defined in \refEq{psf_lms_color_primary}.
To perform the above optimization, we compare images using a loss function (\eg least-squared error) to calculate the erorr between the ground truth image and the retinal image, $\mathcal{L}(r(x, y, c), t(x, y, c))$, where $x$ and $y$ represent image coordinates and $c$ the color channels, which could be in RGB or LMS color opponency spaces.
Note that we have also built a learned equivalent of our approach, which we will detail in the \refSec{implementation}.

\section{Implementation}
\label{sec:implementation}
Our approach is comprised of three primary elements: a color perception model, a prescription correction optimization pipeline and a learned model that demonstrates that our differentiable pipeline can be learnt.
All of these components are implemented on PyTorch \cite{paszke2017automatic}.

\begin{figure}[!ht]
\begin{centering}
\hspace{0.035\textwidth}\begin{minipage}{.43\textwidth}
\begin{lstlisting}[language=IPython,escapeinside={(*}{*)},caption={Computing L, M, S kernel triple for each of the R, G, B channels and create forward model kernel in 4D tensor form. The abstraction is Pythonic.}, label=list:4D_LMS_kernel]
import torch

def get_LMS_kernel(spectrum):
  [(*\textbf{$\lambda_R, \lambda_G, \lambda_B$}*)] = get_display_spectrum()
  for (*\textbf{$i$}*), (*\textbf{$\lambda_{i}$}*) in enumerate([(*\textbf{$\lambda_R, \lambda_G, \lambda_B$}*)]):
      wavelengths=torch.arange(400, 701)
      psf = generate_psf((*\textbf{$\lambda_{i}$}*))
      weighted_psf = spectrum[(*\textbf{$\lambda_{i}$}*)] * psf
      LMS_kernel += convert_to_lms(weighted_psf, (*\textbf{$\lambda_{i}$}*), spectrum[(*\textbf{$\lambda_{i}$}*)])            
  return LMS_kernel

kernel = torch.zeros(3, (*\textbf{$H$}*), (*\textbf{$W$}*), 3)
kernel[0] = [(*\textbf{$H$}*), (*\textbf{$W$}*), get_LMS_kernel(*\textbf{$(red_{spectrum})$}*)]
kernel[1] = [(*\textbf{$H$}*), (*\textbf{$W$}*), get_LMS_kernel(*\textbf{$(green_{spectrum})$}*)]
kernel[2] = [(*\textbf{$H$}*), (*\textbf{$W$}*), get_LMS_kernel(*\textbf{$(blue_{spectrum})$}*)]


\end{lstlisting}
\end{minipage}
\end{centering}
\end{figure}

\begin{figure*}[!ht]
\begin{centering}
\includegraphics[width=2.0\columnwidth]{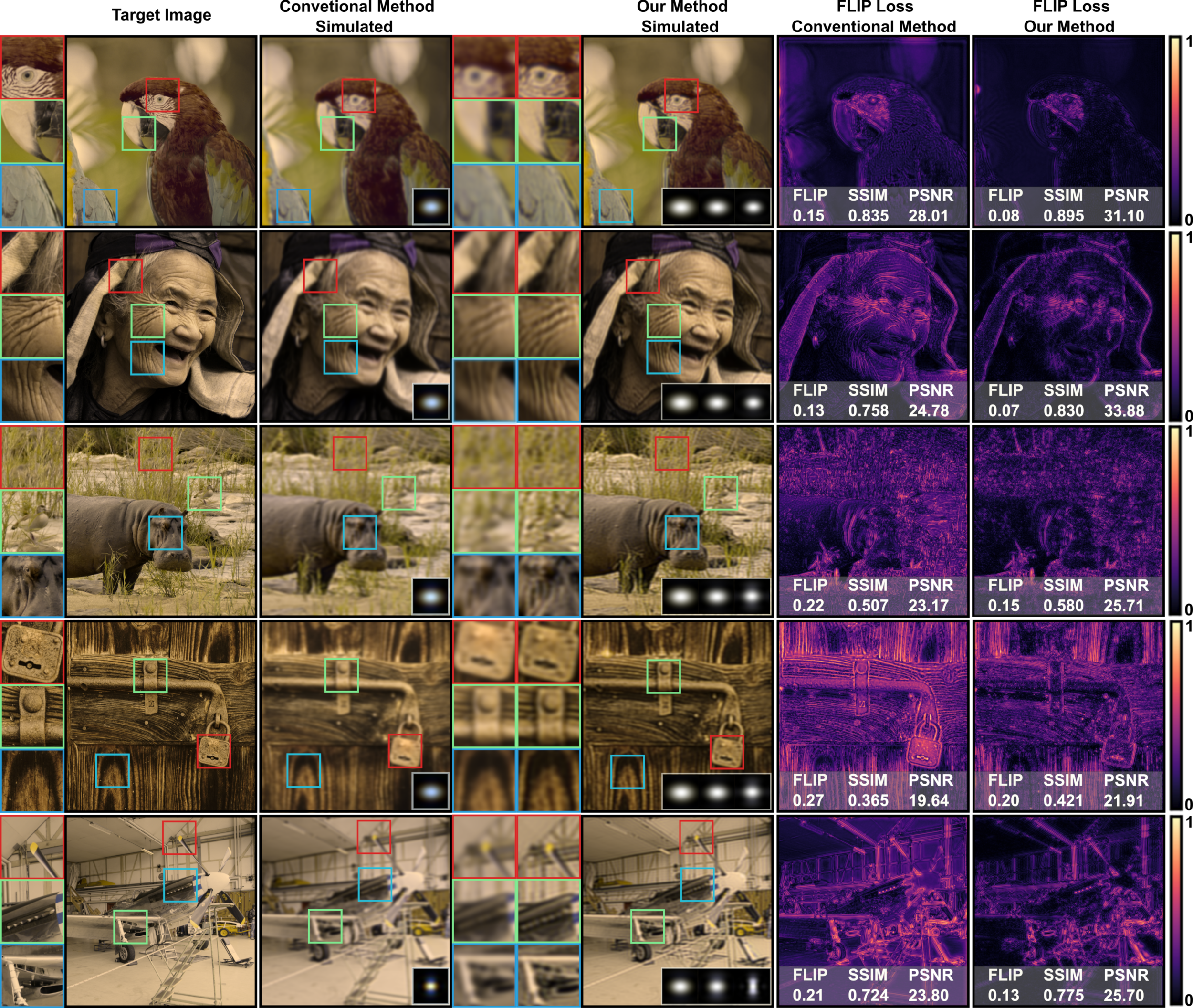}
\caption{
Here we compare outputs from five different refractive vision problems (myopia, hyperopia, hyperopic astigmatism, myopic astigmatism, and myopia with hyperopic astigmatism) for five sample input images. 
We provide simulated LMS space representations of target image, conventional method output, and our method. 
FLIP per-pixel difference along with it's mean value (lower is better), SSIM and PSNR are provided to compare performance of methods. 
Our method shows better loss numbers for each image quailty metrics for each experiment in simulated LMS space.
The contrast improvement by using our method against conventional method also can be obvserved perceptually. 
Source images are from DIV2K image dataset \cite{Agustsson_2017_CVPR_Workshops}.
}
\label{fig:evaluation_image}
\end{centering}
\end{figure*}

\subsection{Color Perception Model}
\label{subsec:color_perception_model}

Firstly, we identify the emitted wavelengths from the subpixels of a target display device.
For that purpose, we acquire the spectrometer data for a target display consisting of discrete wavelengths and their corresponding intensity values normalized between zero and one.
We use \MLP to fit a curve on this discrete data to achieve a vector representation of our intensity profile of color primaries with respect to wavelength.
Our \MLP has 64 hidden layers and converges over 1000 iterations in training with a learning rate of 0.0005.
Once we have numerically identified the normalized intensity of each color primaries as a function of wavelength, we use these 2D (intensity, wavelength) vectors to create our color perception based kernel in LMS space.
For each color primary, we create the set of \PSF based on our zernike polynomial generator by sampling wavelengths from 400 to 700 with 1 nm intervals.
During each sampling step, we create weighted kernels by multiplying the created \PSF with the intensity value based on corresponding wavelength from our created 2D vectors for each colopr primary.
After creating the weighted kernel in each sampling step, we obtain LMS cone responses of weighted kernel using the same intensity and wavelength data. 
To compute LMS cone responses, we use the method explained in section 3.1.
In the last step, the set of weighted kernels are summed up to create our color perception based kernel for each color primary forming a 4D tensor as \emph{[Color Primary, H, W, LMS Response]}.
Our method differs from the conventional method both in terms of kernel type, and convolution operation. 

In conventional method, kernel is a 3D tensor with RGB channels while in our method we use 4D tensor. 
In this 4D tensor formed kernel, each color primary has LMS triple seperately as  \emph{[3, H, W, 3]}.
The LMS based kernel convolves the image's each color channel with corresponding each display spectrum LMS responses. 
This operation computationally more expensive compared to conventional method, since more matrix operation is needed. 
We provide a pseudo-code for constructing our LMS based kernel as in Listings \ref{list:4D_LMS_kernel}.

\subsection{Optimization Pipeline}
\label{subsec:optimization_pipeline}
We implement a prescription correction optimization pipeline using a modern machine learning library with automatic
differentiation \cite{paszke2017automatic}.
Source code of our implementation is publicly available at \codeoptimization \cite{prescription_optimization2022} and \codemodel \cite{learned_prescription2022}.

\text{\emph{Optimization loop:}}
The differentiable input RGB image initialized from the our target RGB image, and it is passed through the forward model during optimization loop.
In forward model, each color channel of initialized input RGB image convolved with the LMS kernel created in computatinal color pipeline.
For example, red channel of input RGB image is convolved with L, M, S channel of red spectrum kernel in LMS space. 
Other color channels of input RGB image are convolved with the same method.
The resulting simulated image represents the image formed on the retina from L, M, S cone activations.
The target image is converted to LMS space to calculate L2 loss against the simulated image in LMS space, which is back-propagated through the optimization model to the input RGB image.
Our results are obtained using Stochastic Gradient Descent with ADAM \cite{Kingma2014Dec} as the optimizer.
Our pipeline is available to be used in NVIDIA GPU accelerated computer.

\subsection{Learned Model}
\label{subsec:unet_training}
We implement a semi-supervised deep learning model capable of reconstructing optimized images from their original RGB versions.
We use a U-Net architecure \cite{Ronneberger:2015} for this purpose.
Such solution is more suitable than an iterative process for achieving real-time applications.
But it trades the image quality for a faster rendering speed.
Our model comprises of 2 outer layers linked to 8 convolutional hidden layers symmetrically connected by skip connections.
Each layer on the contractive path of the model are formed by a double convolution and a max pooling operation. 
On the expanding path, an up-sampling operation with bilinear interpolation initiates each convolution.
During training, batch normalization and ReLU activation are used.

Our model was evaluated on a machine with an NVIDIA GeForce RTX 2070 GPU.
The training dataset comprises of 20 images of dimension 512 x 512 pixels, the RGB images were obtained from Zhang \etal's color image processing dataset \cite{Color_image_processing_dataset} and the target optimized images were generated using our iterative method.
A learning rate of $ 1 \times 10^{-4}$ was used for the training phase and a conventional mean-squared-error loss function guides the stochastic gradient descent optimization.
With convolutional kernels of size 3x3, each input image sees its channels expand from 3 to 92 and all the way up to 1472 at the latent space.
The results in Figure \ref{fig:learned_model_evaluation} shows the comparison of the corrected image between our original pipeline and the neural network's prediction after over 800 epochs of training.
The average time to generate a single corrected image is 0.0029 seconds with the model as opposed to 8.127 seconds using the original method, a tremendous decrease.

\section{Evaluation}
\label{sec:evaluation}

We divided our experiments in to two sections.
In the first part, we use real hardware to test our methods for defocus prescription.
We used Oculus Quest 1 virtual reality headset, and we placed a defocus lens to create artificial prescription for a camera shot. 
In our experiments we use fixed pose, focus camera to capture images to demonstrate the method's performance.

\begin{figure}[!ht]
\begin{centering}
\includegraphics[width=1.0\columnwidth]{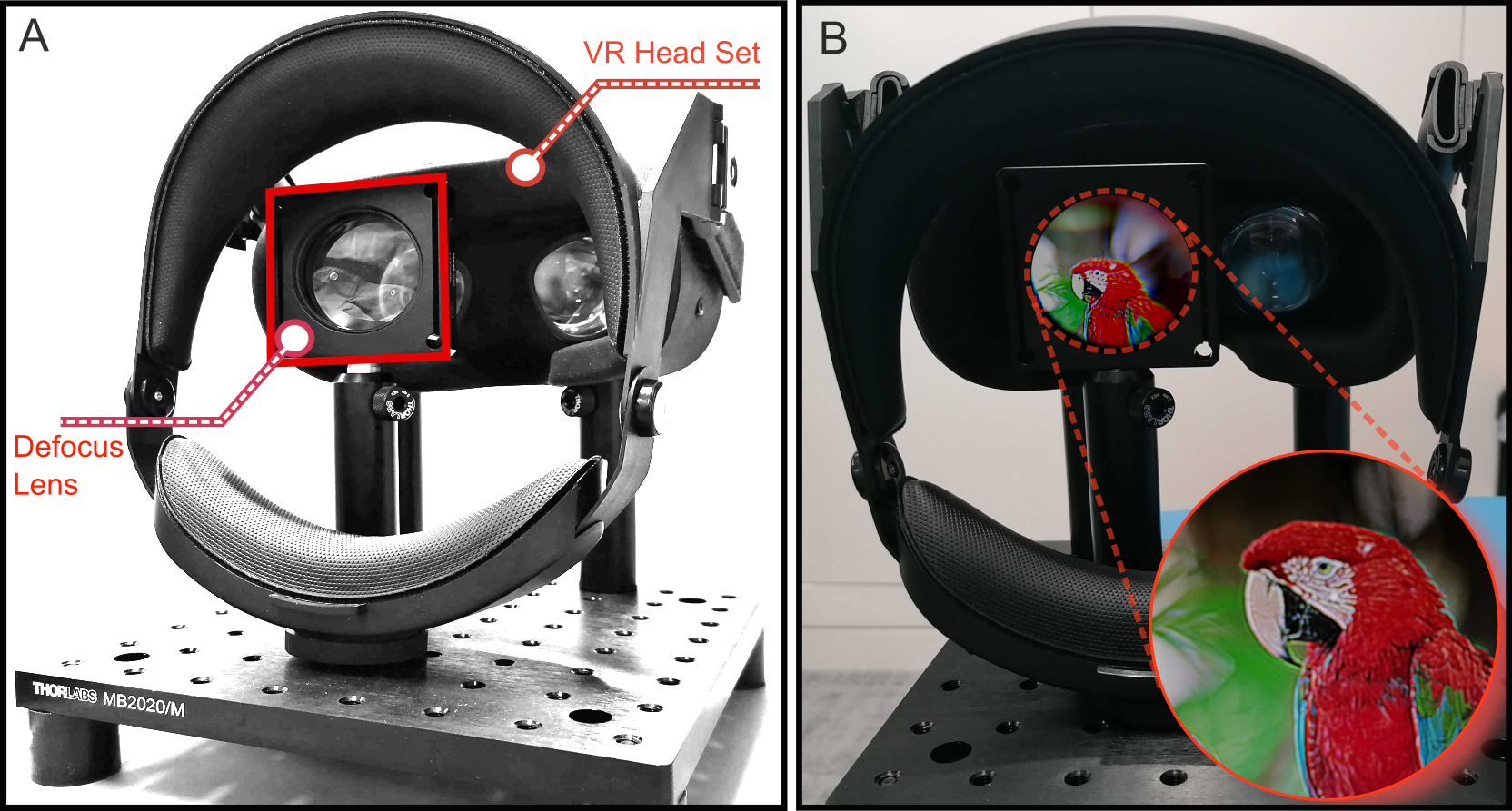}
\caption{
Testbed used in our evaluations.
(A) We use a virtual reality headset and a camera to capture images from our virtual reality headset.
To emulate a prescription problem in the visual system, we use a defocus lens. 
(B) We take pictures with fixed pose and camera focus from behind the defocus lens to evaluate reconstructed images. 
}
\label{fig:vr_setup}
\end{centering}
\end{figure}

\begin{figure}[!ht]
  \begin{centering}
  \includegraphics[width=1.0\columnwidth]{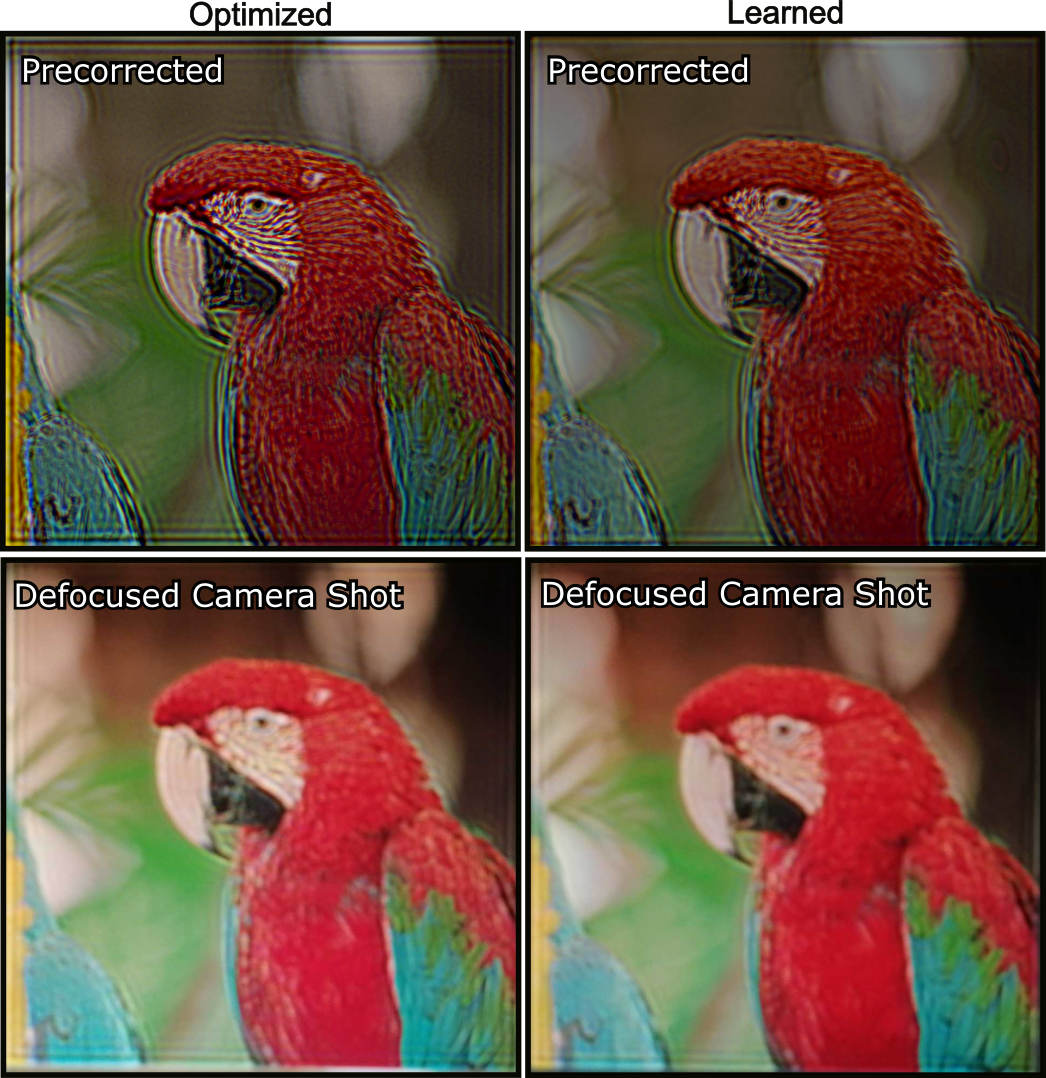}
  \caption{
    Results from our learned model. We compare our optimization pipeline against our learned model.
    The top row shows precorrected images reconstructed by our optimizer and learned model.
    The bottom row shows photographs for each case when captured with a defocused camera.
  }
  \label{fig:learned_model_evaluation}
  \end{centering}
  \end{figure}

Figure \ref{fig:vr_setup} shows our experimental setup for defocus experiments. 
Figure \ref{fig:teaser} shows results from the first part of our experiments.
We modeled the myopia defocus, since in this way we can use defocus lenses to replicate eye prescription. 
Experiments shows that we improved contrast and color compared to conventional method. 
In fact, our method is not able to produce same quailty with the target image.

In the second part, we evaluated our method with different prescriptions to model different refractive eye problems in simulated retinal image representation.
Thus, all the images used in this part are evaluated in simulated LMS space. 
Selected images are aimed to have both high frequency and low frequency features.
Four common different prescriptions are chosen which are myopia, hyperopia, myopic astigmatism, hyperopic astigmatism to test our method against the conventional model.
Also, we tested our method for myopia with hyperopic astigmatism as a complicated refractive eye problem which is not trivial for eyeglass correction.
In each refractive eye problem modelling, +/-1.5D refractive error  is used to model prescriptions.
Resutls are visualized in Figure \ref{fig:evaluation_image}.
We use different image quality measures to compare our method agains the conventional method. 
Our primary chosen image quailty metric is FLIP which compares the images by using principles of human perception \cite{Andersson:2020}.
FLIP allows per-pixel difference loss maps in magma color which is used in our evaluation images to visualize difference in each pixel against the ground truth image. 
In each pixel comparison FLIP counts both color and edge differences based on models of \HVS. 
Therefore, we believe that this metric fits with our work. 
Although many research on this area has been used SSIM or PSNR loss, however FLIP is adventegous as it is adhering human visual system while others are not \cite{Nilsson:2020}.
In addition to FLIP, we use SSIM and PSNR to compare our method agains to conventional method to be stayed relevant with the research community. 
Figure \ref{fig:evaluation_image} demonstrates the comparison of our method against the naive method with our perceptually guided color modelling.

Results shows that color opponency based kernel modelling improves the contrast of retinal output image.
Selected 3 areas are mangified to show visibility of improvements in a detailed way.
Fom the per-pixel difference loss maps, we found that our method is better in low frequency features while our method provides slight improvement in high frequency parts of images.
In overall, it is shown that perceptually guided color based kernel has better contrast compared to conventional method.

\section{Discussion}
\label{sec:discussion}
Our method could be potentially integrated with Mandl \etal \cite{Mandl2021Oct} to support a broader user base with refractive vision impairments.
To the best of our knowledge, we provide encouraging results improving the conventional method in the literature.

\paragraph{Spatially Varying \PSF.}
Our method does not account for spatially varying natures of \PSF in the \HVS, which often arrives with computational cost and complexity \cite{heide2013high}.
We designed our implementation in constant resolution displays instead of varying resolution ones like foveated displays.
As an alternative, the deep learning methods can help support spatially varying \PSF convolutions in the modeling \cite{Yanny:22} with lesser computational cost but with demand in data for training.
Thus, our method can benefit from these techniques in the future for precision modeling.

\paragraph{Chromatic Aberrations In A Human Eye.}
We use \PSF created by the same Zernike coefficients for each wavelength in our forward model.
However, optics of \HVS contain chromatic aberrations that are wavelength-dependent.
As a future work, we can further improve the accuracy of our modeling for a human observer by taking into account the chromatic aberrations in the \HVS.
In the meantime, curious readers can find greater details regarding chromatic aberrations in work by Cholewiak \etal  \cite{ChromaBlur2017}.

\paragraph{Image Quality.}
Approaches for prescription correction with additive displays are fundamentally limited.
This limit stems from the fact that \PSF, the non-negative transfer function of an additive display, could support a limited range of frequencies and cause contrast loss.
Our work could be made to be complementary to holographic displays \cite{aydindougan2021applications, chakravarthula2021gaze, walton2022metameric}, which promise a unique solution for this issue originating from non-negativity in additive displays.

\begin{figure}[!ht]
\begin{centering}
\includegraphics[width=1.0\columnwidth]{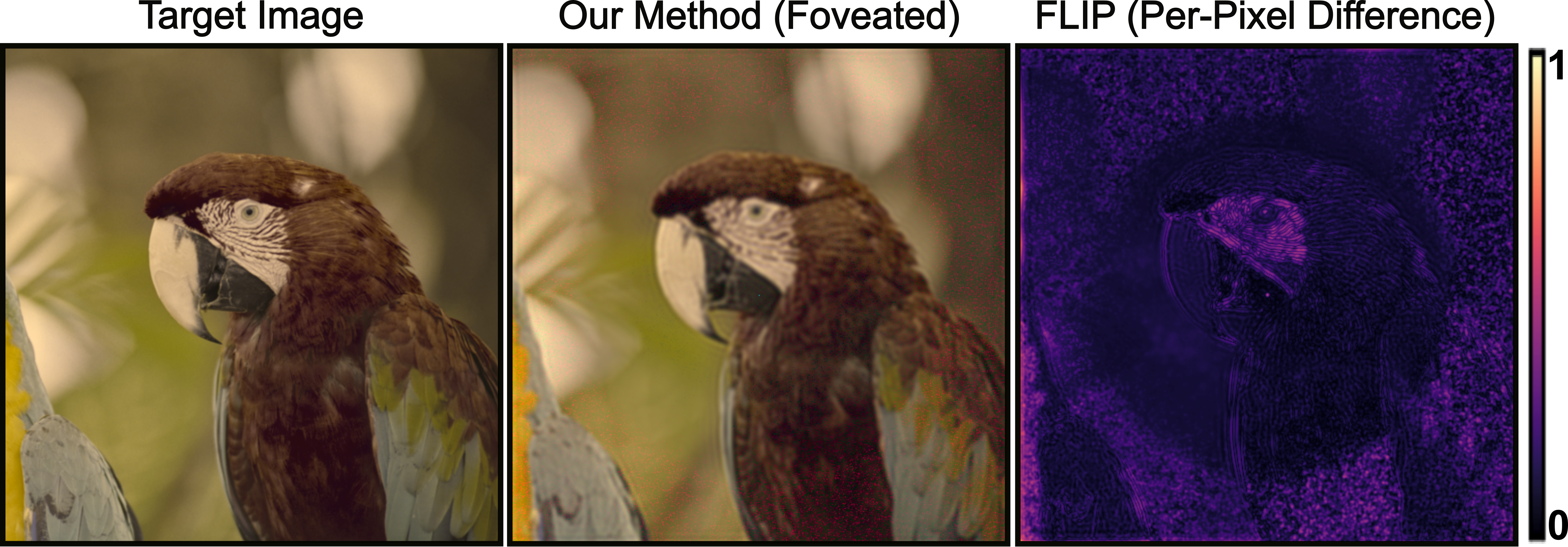}
\caption{
We reconstructed image in our method with addition of foveation.
Foveated rendered area is in the center of reconstructed image. 
FLIP per-pixel difference map highlights the foveation.
}
\label{fig:parrot_discussion_figure}
\end{centering}
\end{figure}

\paragraph{Foveated Rendering.}
Foveated rendering in graphics \cite{walton2021beyond} and displays \cite{kim2019foveated} has garnered interest in the \VR and \AR research community.
We believe that our method can also benefit from this trend by accounting for trends in chromatic and achromatic contrast sensitivity \cite{venkataraman2019peripheral, williams1993color, hansen2009color} in the \HVS.
Moreover, we could add a rod's response to cone responses by reformulating the LMS response to improve color difference predictions \cite{ashraf2022change}.
We will explore this path in our future work (See Figure \ref{fig:parrot_discussion_figure} for our early results.)

\section{Conclusion}
\label{sec:conclusion}
Identifying means to help display users with their vision impairments is an essential aspect of graphics systems.
As we focus on this critical issue, we present a new rendering approach that provides sharp images when viewed by users with vision impairments without their prescription glasses.
Specifically, our rendering approach uniquely merged key insights from \HVS. 
It showed that it could help improve visual experiences and comfort in \VR headsets by enhancing color and contrast in the displayed images.
The future will likely bring more principled approaches in \AR/\VR displays (\eg holographic displays), which could enable future research investigations based on findings from this work.

\acknowledgments{
The authors would like to thank Praveen Selvaraj for the engineering support through early phases of this project.
Kaan Ak\c{s}it is supported by Meta Reality Labs inclusive rendering initiative for building the rendering pipeline.
}

\bibliographystyle{abbrv-doi}
\bibliography{references}



\section*{Supplementary material}

# ChromaCorrect: Prescription Correction in Virtual Reality Headsets through Perceptual Guidance: Supplementary Material

Ahmet Güzel*  
University of Leeds

Jeanne Beyazian†  
University College London

Praneeth Chakravarthula‡  
UNC Chapel Hill  
Princeton University

Kaan Akşit§  
University College London

**Index Terms:** Human-centered computing—Visualization—Visualization techniques—Treemaps; Human-centered computing—Visualization—Visualization design and evaluation methods

## 1 Contributions

Here we list contributions from each author in this specific research work:

Ahmet Güzel:

- Coding an entire computational pipeline for color model and optimization.
- Compiling extensive study on evaluation.
- Compiling majority of the figures.
- Writing, review and edits for the entire documentation.

Jeanne Beyazian:

- Compiling a dataset and development and training of machine learning model.
- Helping with the figures.
- Writing sections related to machine learning.

Praneeth Chakravarthula:

- Mentorship (Ahmet Güzel).
- Steering the project towards Virtual Reality applications.
- Review and edits for the entire documentation.

Kaan Akşit:

- Initiation of project idea, this includes blueprints for projector and algoritmic designs, and early experimentations.
- Literature review and mentorship (Ahmet Güzel and Jeanne Beyazian).
- Writing, review and edits for the entire documentation.

*e-mail: od20ahg@leeds.ac.uk  
†e-mail:jeanne.beyazian.21@ucl.ac.uk  
‡e-mail: praneethc@princeton.edu  
§e-mail: k.aksit@ucl.ac.uk

## 2 Color Opponency

Human retinal cells could be broadly classified into rods and cones. Cone cells are primarily responsible for color perception in Human Visual System (HVS), and the cone density on our retinas peek at our fovea and drops sharply towards larger eccentricities [4]. Cone cells have three subtypes known as Short (S), Medium (M), and Long (L) cells, where each differs in sensitivity to wavelengths of light [6, 7]. These L, M, and S cones reduce wavelengths of incoming light into trichromat values by integrating them over their response functions [8]. According to earlier color opponency studies, HVS relies on comparison of L, M and S cone activation outputs [3]. A widely used modeling method in color opponency is S versus (M+L) for blue-yellow (BY) channel, and L versus M for red-green (RG) channel [2]. Schmidt *et al.* [5] proposes L versus (M+S) opponency for BY channel and M versus (L+S) opponency for RG channel.

## 3 Learned Model

Figure 1 shows further results for comparison of the corrected image between our open-box model and the neural network's prediction after over 800 epochs of training.

## 4 Futher Evaluation

Figure 2 shows further results for different myopia starting from -2.0 dioptres to -4.5 dioptres. We show that our method's reconstructed images have better constrast than conventional method's reconstruced images for each case. Images are simulated in LMS space.

Figure 3 shows further results for different images from DIV2K image dataset [1]. In each refractive eye problem modelling, +/-1.5D refractive error is used to model prescriptions.

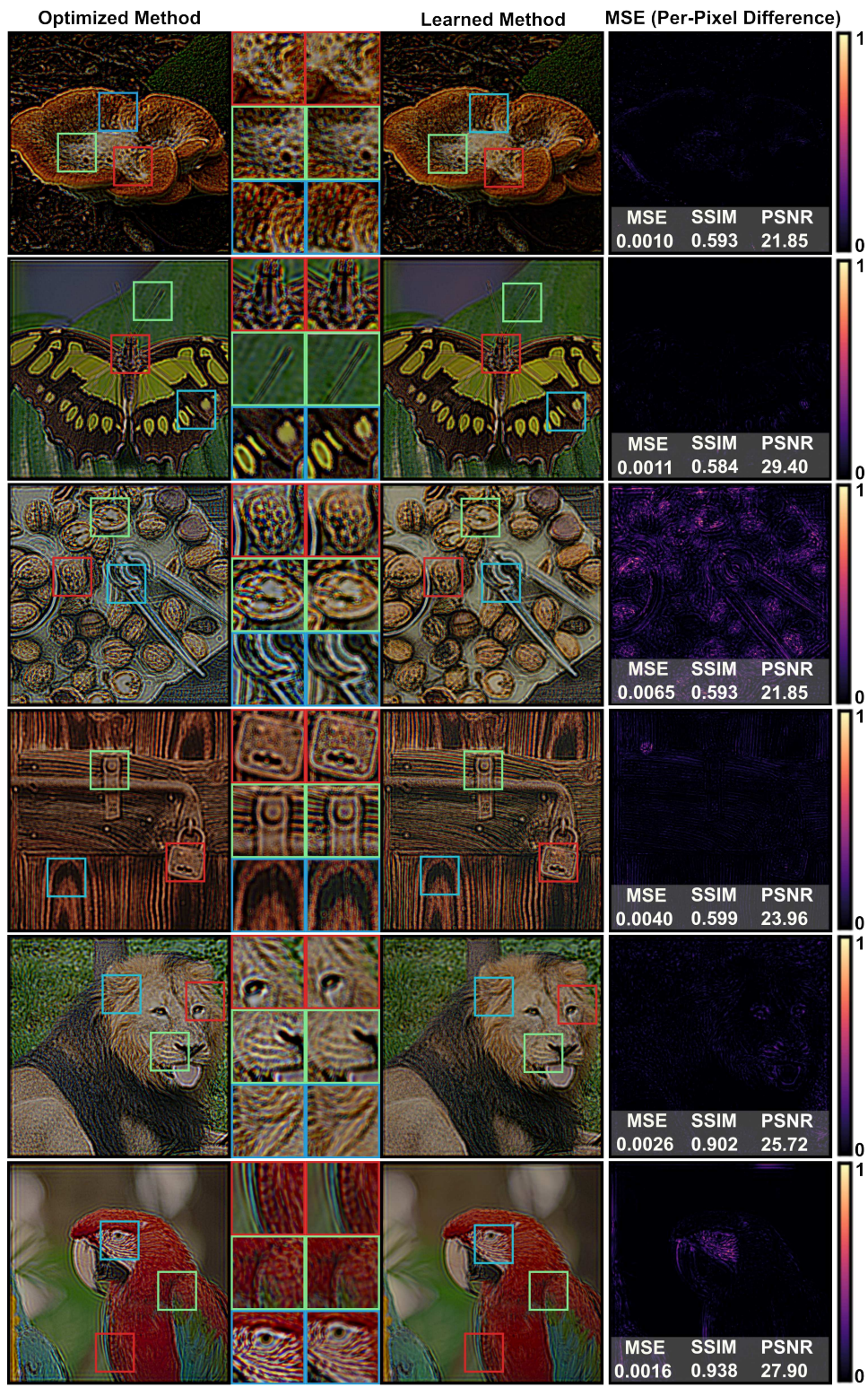

Figure 1: Additional images for learned method. Images are selected from DIV2K image dataset [1].

Figure 2: Comparison of convetional method and our approach with different myopia cases. Images are simulated in LMS space.

| Target Image | Conventional Method | Our Method |
|---|---|---|
| 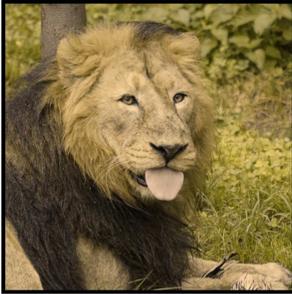 | 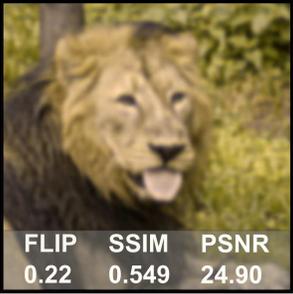 FLIP 0.22  SSIM 0.549  PSNR 24.90 | 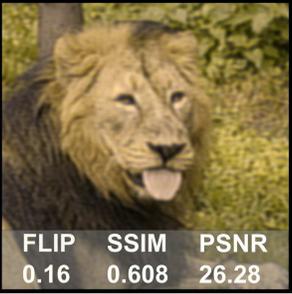 FLIP 0.16  SSIM 0.608  PSNR 26.28 |
| 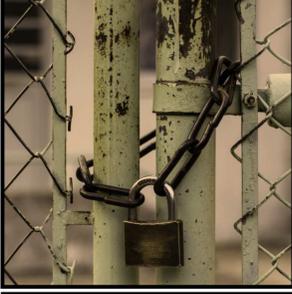 | 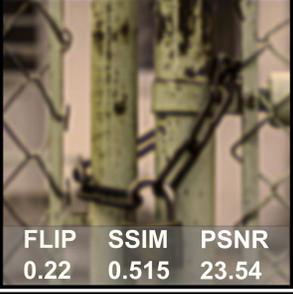 FLIP 0.22  SSIM 0.515  PSNR 23.54 | 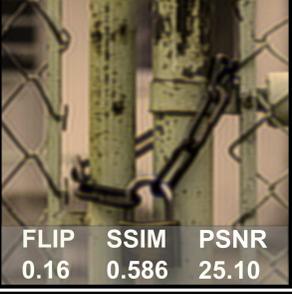 FLIP 0.16  SSIM 0.586  PSNR 25.10 |
| 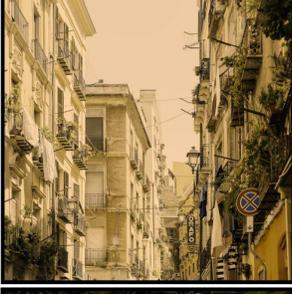 | 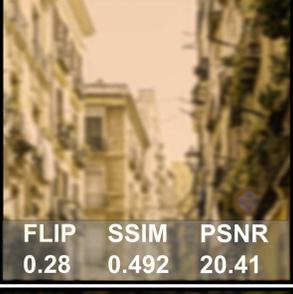 FLIP 0.28  SSIM 0.492  PSNR 20.41 | 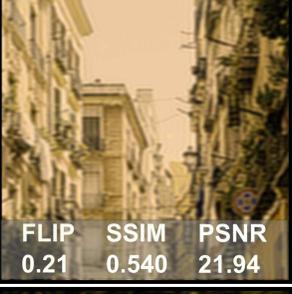 FLIP 0.21  SSIM 0.540  PSNR 21.94 |
| 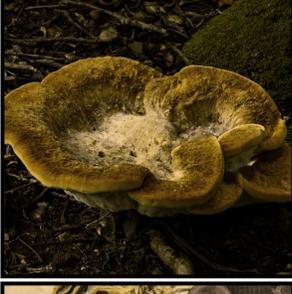 | 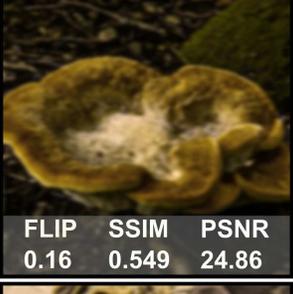 FLIP 0.16  SSIM 0.549  PSNR 24.86 | 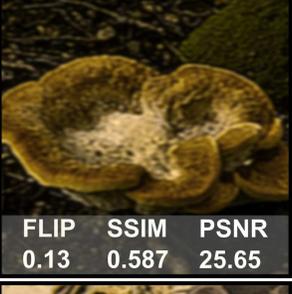 FLIP 0.13  SSIM 0.587  PSNR 25.65 |
| 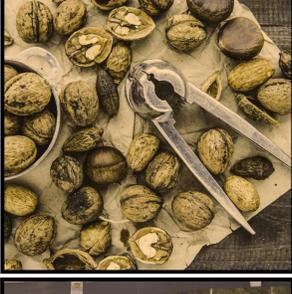 | 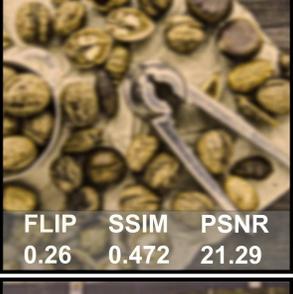 FLIP 0.26  SSIM 0.472  PSNR 21.29 | 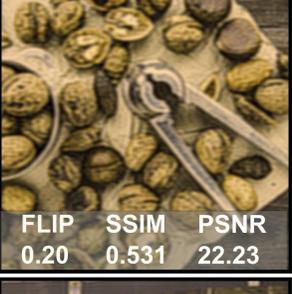 FLIP 0.20  SSIM 0.531  PSNR 22.23 |
| 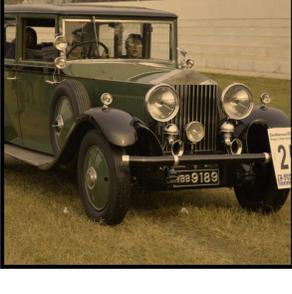 | 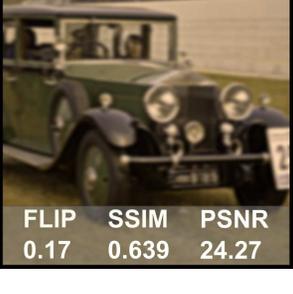 FLIP 0.17  SSIM 0.639  PSNR 24.27 | 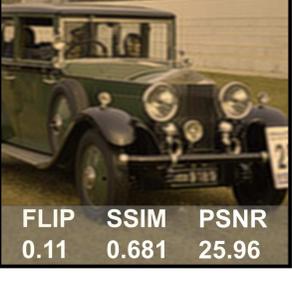 FLIP 0.11  SSIM 0.681  PSNR 25.96 |

Figure 3: Additional images for the evaluation section. Images are simulated in LMS space.


\end{document}